# Multiple Quantum Well AlGaAs Nanowires

Chen Chen[1], Nadi Braidy[2], Christophe Couteau[4],
Cécile Fradin[3], Gregor Weihs[4], Ray LaPierre[1*]

[1]Centre for Emerging Device Technologies, and Department of Engineering Physics,
McMaster University, Hamilton, Ontario, L8S 4L7, Canada

[2]Department of Materials Science and Engineering,
McMaster University, Hamilton, Ontario, L8S 4L7, Canada

[3]Department of Physics and Astronomy,
and Department of Biochemistry and Biomedical Sciences
McMaster University, Hamilton, Ontario, L8S 4M1, Canada

[4]Institute for Quantum Computing, University of Waterloo
200 University Avenue West, Waterloo, Ontario, N2L 3G1, Canada

*Corresponding author. Email: lapierr@mcmaster.ca

**Abstract**

This letter reports on the growth, structure and luminescent properties of individual multiple quantum well (MQW) AlGaAs nanowires (NWs). The composition modulations (MQWs) are obtained by alternating the elemental flux of Al and Ga during the molecular beam epitaxy growth of the AlGaAs wire on GaAs (111)B substrates. Transmission electron microscopy and energy dispersive X-ray spectroscopy performed on individual NWs are consistent with a configuration composed of conical segments stacked along the NW axis. Micro-photoluminescence measurements and confocal microscopy showed enhanced light emission from the MQW NWs as compared to non-segmented NWs due to carrier confinement and sidewall passivation.

Interest in semiconductor nanowires (NWs) has increased remarkably over the past decade due to their ability to directly form nanometer scale structures. While various methods exist to synthesize semiconductor NWs, the vapor-liquid-solid (VLS) approach where a metal seed particle determines the growth of the NW crystal, has received the most attention. This approach was first reported by Wagner and Ellis more than 40 years ago to explain the epitaxial growth of micron-sized Si whiskers.[1] Interest in this mechanism was renewed in the early 1990s, when Hiruma and co-workers demonstrated its applicability at the nanometer scale to fabricate III-V NWs.[2] Furthermore, NWs may be doped and heterostructures may be formed along the axis or radius of NWs, which are essential features for optoelectronic devices. It is now well recognized within the research community and the semiconductor industry (see for instance the International Technology Roadmap for Semiconductors[3]) that one-dimensional structures will play a major role in the efforts towards device miniaturization and the development of novel applications. Indeed, semiconductor NWs have already demonstrated their potential in a wide range of electronic, photonic, and sensing devices.[4-8] Previously, we reported on the growth of AlGaAs NWs for applications in the visible light spectrum.[9] In the present letter, we extend this previous study by reporting on the growth and characterization of AlGaAs NWs with multiple quantum wells (MQWs) having an improved luminescence efficiency.

Our NWs were grown by the VLS growth process using Au nanoparticles as seeds for site selective growth. Substrates of GaAs (111)B were first submitted to a 20-minute UV-ozone treatment, etched in a 10% buffered HF solution, and rinsed with deionized water. The samples were then transported in ambient air to an e-beam evaporation system, where a 1 nm thick film of Au was deposited. The samples with Au were then transferred in ambient air to a gas source molecular beam epitaxy (GS-MBE) growth chamber. In GS-MBE, group III species (Al and Ga) are supplied as monomers from a heated solid elemental source, and the group V species are supplied as dimers ($As_2$) from a hydride ($AsH_3$) gas cracker operating at 950ºC. Prior to the actual growth, the Au-covered substrates were heated to a temperature of 500°C for 5 minutes under an $As_2$ flux to form Au nanoparticles on the surface. Simultaneous desorption of native oxide was enhanced by the use of an inductively-coupled hydrogen plasma source. Following the oxide removal, the temperature was set to 570°C for NW growth. This process results in Au nanoparticles with diameters ranging between 10 and 40 nm. The growth of heterostructures was achieved by alternating the group III source flux during growth of the NWs. Referring to Figure 1, the NW growth consisted of three AlGaAs quantum wells (W1, W2, W3) separated by AlGaAs barriers of higher Al concentration (B1 to B4). The relative composition of Al and Ga was controlled by the relative flux of Al to Ga set by the effusion cell temperatures with nominal well composition of $Al_{0.22}Ga_{0.78}As$ and nominal barrier composition of $Al_{0.52}Ga_{0.48}As$. The barriers were nominally undoped while the wells were p-doped with Be to $10^{18}$ cm$^{-3}$, as determined by earlier calibration of thin film doping based on GaAs (100) epilayers. This doping was performed for the purpose of other studies that will not be discussed here. Growth occurred with $AsH_3$ flow of 3 sccm and a V to III flux ratio of 2.0. The equivalent two-dimensional growth rate was estimated at 1 µm/hr, as determined from previous thin film calibrations on GaAs (100) substrates. The procedure was initiated by opening the Al and Ga shutters for 10 minutes in order to grow a nominally undoped AlGaAs barrier segment (referred to as B1, in Figure 1). The growth of a p-doped AlGaAs well (W1) then followed for 3 minutes. The latter procedure was repeated twice, to produce wells W2 and W3 and their corresponding barriers. Finally, the NWs were capped with an undoped

AlGaAs barrier using an additional 5 minutes deposition. Each of the well/barrier interfaces was accompanied by a growth interruption of one minute to adjust the Al/Ga cell temperatures to the desired composition.

The growth of NWs by GS-MBE occurs by adatoms impinging on the substrate surface, which subsequently diffuse in a "random walk" manner to the base of the wires, then along the wire sidewalls towards the Au-nanowire interface. Most NWs to date have been grown using metal organic vapour phase epitaxy (MOVPE) using metalorganic group III sources (e.g., trimethylgallium) and group V hydrides.[10] In the case of MOVPE, the NW surface acts as a site for the decomposition of gaseous precursors whereas in the case of MBE the constituent elements arrive at the substrate already decomposed. Thus, VLS growth by MBE includes substantial radial growth of the NWs due to uncatalyzed deposition on the NW sidewalls. Our previous studies have shown[9] that under the growth conditions of the present study, the NWs will experience both axial VLS growth by adatom accumulation in the Au seed particle, as well as radial growth due to non-VLS deposition on the sidewalls of the NWs. As a result, NWs by GS-MBE grow in the core multi-shell geometry illustrated in Figure 1, as will be verified below. The elemental supply and pre-cracking of the elements in GS-MBE also means that growth can take place at relatively low V/III flux ratios (generally in the range of 1.5 to 2). These conditions generally favor more control over NW growth, particularly in the fabrication of compound semiconductor heterostructures, as compared to MOVPE.

After growth, the NWs were removed from the substrate by immersing a cleaved portion of the substrate in methanol and sonicating for 1 to 2 minutes. A small volume (~10 µL) of the NW suspension was left to dry onto a holey carbon film supported by a Cu mesh

TEM grid. Bright-field (BF) and high-angle annular dark-field (HAADF) observations were performed with a JEOL 2010F field-emission gun TEM operated at 200 kV in scanning mode with a ~0.7 nm diameter probe. HAADF micrographs were taken with a Fishione detector with a collection angle of 70 mrad. Energy dispersive X-ray spectroscopy (EDS) measurements were collected with a Si(Li) X-ray ultrathin window energy dispersive spectrometer (Oxford Instruments). Elemental profiles were generated by collecting X-rays with energies corresponding to the Al K$\alpha$, Ga K$\alpha$, As K$\alpha$, and Au L$\alpha$ peaks using the INCA software.

Fluorescent imaging was performed by allowing a drop of the sonicated suspension of NWs to dry on a glass substrate. The deposited NWs were covered with a drop of glycerol oil to assist confocal microscopy performed with a Leica TCS SP5 instrument with a Planapo 63x/1.3 NA objective. Excitation was provided by a 488 nm argon ion laser and detection was performed using a Hamamatsu photomultiplier tube. A prism spectrometer, placed in front of the photomultiplier tube detector, selected a wavelength range of 710 to 770 nm for detection.

NWs in methanol solution were also dispersed onto silicon substrates for micro-photoluminescence ($\mu$PL) measurements in a continuous flow helium cryostat at 10 K. PL excitation and collection were performed through a microscope objective with numerical aperture of 0.7, providing a spot diameter of about 1 $\mu$m. The excitation was provided by a HeNe laser at wavelength of 632 nm and a power of 2.5 $\mu$W. PL was resolved by a 75 cm grating spectrometer, and detected by a liquid nitrogen cooled Si CCD camera.

0.5 to 0.8 $\mu$m-long NWs were observed with a tapered morphology, terminated by a Au nanoparticle. The NWs had a diameter ranging from 80 to 100 nm near the base and between 10 to 40 nm at the tip (Figure 2a). As discussed previously[9], when the wire is short, growth occurs by diffusion of adatoms from the base to the tip of NWs. However, as the height of NWs increases, the diffusion of adatoms to the tip of the NWs becomes increasingly unlikely and sidewall deposition becomes the dominant growth mechanism, thus leading to the tapered morphology. Thin bright bands intersecting the NW perpendicular to the growth direction (Figure 2a) correspond to stacking faults, commonly observed in NWs grown in the [111]B direction[11,12]. Electron diffraction from single NWs (not shown) indicated a zincblende crystal structure and stacking faults consisting of insertions of wurtzite crystal structure. This polyptism results from the small energy difference between zincblende and wurtzite crystal structures[13], and will not be considered further in this letter.

EDS measurements were performed to probe the elemental distribution along the length and the diameter of the NWs. A full quantitative analysis of the EDS signal will not be performed here. Instead, it will suffice for the purpose of this paper to consider that the compositional variation of the individual well and barrier segments will be reflected by the relative intensity of the elemental X-ray signals. The EDS linescan shown in Figure 2(b) was performed along the axis of the NW of Figure 2(a), following the trace of the dotted line. Given the constant stoichiometry of As across the NW, the decrease in the As profile follows the tapering of the NW. More importantly, the Al and Ga counts revealed three regions where the Al and Ga signal variations are anti-correlated. The regions with lower Al signal and correspondingly higher Ga signal are indicative of the presence of the three well segments, namely W1, W2, and W3 (indicated by the double-ended arrows), each having a length of about 100 nm.

To verify the core-shell structure suggested in Figure 1, the NWs were probed in high angle annular dark field (HAADF) STEM. For collection angles larger than ~50 mrad, the contrast becomes much more sensitive to thickness and composition as compared to diffraction-related effects. Since the thickness varies continuously across the wire length and diameter, any abrupt modulations in the HAADF signal can be assigned to compositional fluctuations. The contrast arising from the stacking faults in Figure 3a is much weaker but not completely suppressed, even at a collection angle of 70 mrad. However, the dark bands in Figure 3a running almost parallel to the NW wall can be assigned to the presence of lower average atomic number, i.e. Al-rich zones of the wire. This assignment could be confirmed by EDS linescans performed along the diameter of the NW. Once the position of the well segments along the NW axis are located using an axial linescan as exemplified in Figure 2(b), radial EDS linescans were performed across the top well W3 (upper dashed line in Figure 3(a)) and across the bottom well W1 (lower dashed line in Figure 3(a)), as shown in Figures 3(b) and (c), respectively. The positions of these radial linescans are also indicated by dashed lines in Figure 1. The EDS linescan in Figure 3(b), across W3, agrees with a single core-shell structure with a higher Al concentration on the outer layer of the NW. These correspond to well W3 and barrier B4, respectively, near the top of the NW, as illustrated by the sketch in Figure 1. Similarly, the EDS signal of Al and Ga in Figure 3(c) across W1 are indicative of three Al-poor shells, consistent with the W1, W2, and W3 regions. From the EDS profiles, the thickness of the AlGaAs shells were estimated to be between 5 and 10 nm, which is the range where quantum confinement effects are expected, as will be discussed below.

It is readily apparent that the EDS linescan results may be explained by the core-multishell geometry presented in Figure 1. From the bottom up, the Al-rich (B1) barrier segment was grown followed by the first Al-poor well (W1). However, in addition to the axial growth seeded by the Au particle, there also exists simultaneously some radial growth on the sidewalls of the NWs resulting in a W1 shell. This process is subsequently repeated for each segment of the NW resulting in the observed core-multishell structure. It is important to note that the core-multishell structure was only discernible for NWs having a Au particle below approximately 30 nm in diameter. For NWs seeded from larger Au particles, only the axial wells were detected and the core-shell structure in STEM images and EDS measurements (not shown) was not apparent. This may be explained by the dependence of NW growth rate on Au particle diameter. As reported previously[9,14], the growth rate of NWs decreases with increasing Au particle diameter. This implies that NWs with relatively small Au particles quickly reach a NW height which exceeds the adatom diffusion length. Subsequently, adatoms are unable to reach the tip of NWs and will instead attach to the NW sidewalls and contribute to the core-multishell geometry. Conversely, larger Au particles catalyzed short NWs with heights smaller than the adatom diffusion length, thus favoring axial rather than radial growth. Hence, only the smallest Au particle diameters produced clear core-shell structures.

Next, we investigated the influence of the core-shell geometry on the optical properties of the NWs. First, a Leica confocal microscope was used to image the photoluminescence (PL) emission from NWs prepared on glass substrates. Figure 4(a) reveals the optical image and Figure 4(b) shows the corresponding PL emission for the MQW NWs at room temperature which exhibits strong emission intensity. We compare these results with previously reported[9] $Al_{0.37}Ga_{0.63}As$ NWs grown with a nominally homogeneous composition (without MQWs) and measured under the identical confocal microscopy conditions (Figures 4(c) and (d)). In comparison to the MQW NWs, only a relatively weak emission was observed from a few of the non-segmented NWs (highlighted by circles). The improved luminescence intensity from the MQW NWs can probably be explained by the core-shell structure, which is expected to passivate the NWs, reducing non-radiative surface states on the sidewalls of the NW core.

To further quantify the luminescent properties of the MQW NWs, µPL spectroscopy was performed at 10 K for NWs sonicated onto Si substrates. Results are presented for two representative MQW NWs labeled 1 and 2 in the SEM image of Figure 5(a), corresponding to NWs with Au particle diameters of 32 and 10 nm, respectively, as measured directly from the SEM images. First, the spectrum in Figure 5(b) from NW 1 revealed a dominant peak at about 1.65 eV. This spectrum was typical for NWs with diameters greater than 30 nm. As discussed above, core-shell structures were not observed for NWs catalyzed by a Au particle with a diameter greater than about 30 nm. Only the axial composition modulations with well thicknesses of ~100 nm were detected in such NWs. Quantum confinement effects are unlikely to be a significant contributing factor to the PL results with this configuration since the diameter of the NWs and thickness of the well segments are greater than the exciton Bohr radius, estimated to be 18 nm.[15] We therefore assume that the feature of the PL spectrum in Figure 5(b) from NW 1 corresponds to unresolved exciton-related transitions from the axial wells of the NW. According to the composition dependence of the bandgap and accounting for the peak energy shift with temperature for bulk $Al_xGa_{1-x}As$,[16] the peak at 1.65 eV corresponds to an average Al well composition of $x = 0.10$. This is somewhat lower than the target Al composition of $x = 0.22$ expected from the two-dimensional film calibrations. However, this difference between film and NW composition is expected due to the different growth mechanisms of the NWs compared to planar films. Similar observations have been reported elsewhere for AlGaAs NWs on GaAs substrates.[9,17] A lower composition is expected in NWs where growth is dependent upon the diffusive transport of adatoms along the NW sidewalls towards the Au-NW interface, unlike the growth by direct impingement in the case of 2-D films.

For the NW with smaller Au particle diameter of 10 nm, three higher energy peaks at 1.68, 1.70, and 1.72 eV dominate the spectrum as shown in Figure 5(c) in addition to the feature near 1.65 eV. This spectrum is typical for NWs with diameters below 20 nm where core-shell structures were clearly resolved in HAADF and EDS measurements. The narrowest linewidth observed for these peaks was ~ 1 meV, highly indicative of quantum confinement. A weaker intensity peak near 1.65 eV is probably the bulk-like peak associated with the axial wells as observed in Figure 5(b). The emergence of the high energy PL peaks was concurrent with the emergence of the core multi-shell structure in the smaller diameter NWs. We therefore assume that the sharp lines are exciton-related peaks from the three MQW shells W1, W2, and W3 near the NW sidewalls illustrated in Figure 1. The sharp peaks were not observed in NWs grown without QWs.[9] Although the precise thicknesses and compositions of the radial shells in the MQW NWs are difficult to extract from the encapsulating material background, we may surmise that the different energies of each of the three sharp peaks may be due to different quantum confinement due to slightly varying thickness or composition for each of the three individual shells. When compared to the bulk composition peak of 1.65 eV, the sharp lines correspond to quantum confinement energies ranging from 30 to 70 meV. Using a finite quantum well model, and assuming a nominal barrier composition of $Al_{0.52}Ga_{0.48}As$ and well composition of $Al_{0.1}Ga_{0.9}As$, these confinement energies can be assigned to shell thicknesses ranging from 4 to 7 nm.[18] This thickness range is consistent with that estimated from EDS in Figure 3(c). The peak PL intensity (not shown) reported previously[9] for non-segmented NWs was lower by a factor of

~100 compared to that for the MQW NWs in Figures 5(b). The higher luminescence described earlier in Figure 4 may be due to the carrier confinement of the MQWs in addition to the passivating nature of the core-shell structure.

In conclusion, MQW AlGaAs NWs were grown in a molecular beam epitaxy system on GaAs (111)B substrates. Micro-photoluminescence measurements and energy dispersive X-ray spectroscopy indicated a core-shell structure and axial multiple quantum well structure, producing carrier confinement both in the radial and axial directions. Confocal microscopy showed an improvement in luminescence efficiency compared to non-segmented NWs, and micro-photoluminescence spectroscopy suggested the presence of quantum confinement within the core-shell geometry of the NWs. This work illustrates the relevance of core-shell structures for achieving efficient luminescence from NWs.

**Acknowledgements**

This work was supported by a Nano Innovation grant. The authors thank Brad Robinson and the staff of the CEDT for the GS-MBE growths and the enlightening discussions, and Fred Pearson for assistance with the STEM.

**Figure Captions**

Figure 1. Sketch of the proposed nanowire configuration, as shown in cross-section showing wells W1, W2, and W3 (shaded regions) separated by barriers B1 to B4 (white regions). The dashed lines indicate the location of radial line scans shown in Figure 3.

Figure 2. (a) Bright field scanning transmission electron microscopy micrograph of a nanowire. The dashed line indicates the location of the energy-dispersive X-ray spectroscopy linescan in (b). The Au nanoparticle is visible at the tip of the NW on the right-hand side.

Figure 3. (a)Annular dark field scanning transmission electron microscopy micrograph of a nanowire. The dashed lines on the micrograph indicate the trace of the energy-dispersive X-ray spectroscopy linescans taken across the NW (perpendicular to the growth direction) shown in (b) and (c). Scale bar is 80 nm. Arrows in (b) and (c) indicate the Al deficient shells.

Figure 4. Confocal microscopy image for multiple quantum well NWs (a) and non-segmented NWs (c). (b) PL image corresponding to (a). (d) PL image corresponding to (c). Circles in (d) indicate weak PL emission from two NWs. Scale bar is 3 µm.

Figure 5. (a) SEM image of several MQW NWs. Scale bar is 1 µm. Corresponding µPL spectra are shown in (b) and (c) for the NWs labelled 1 and 2, respectively, in (a).

Figure 1

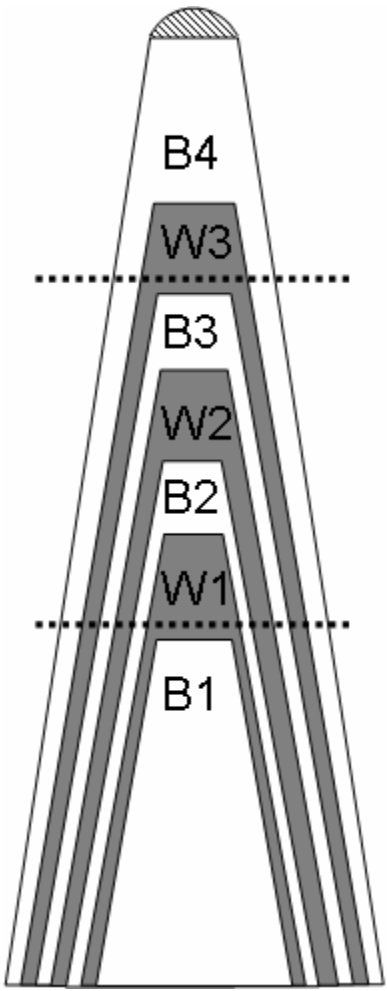

Figure 2

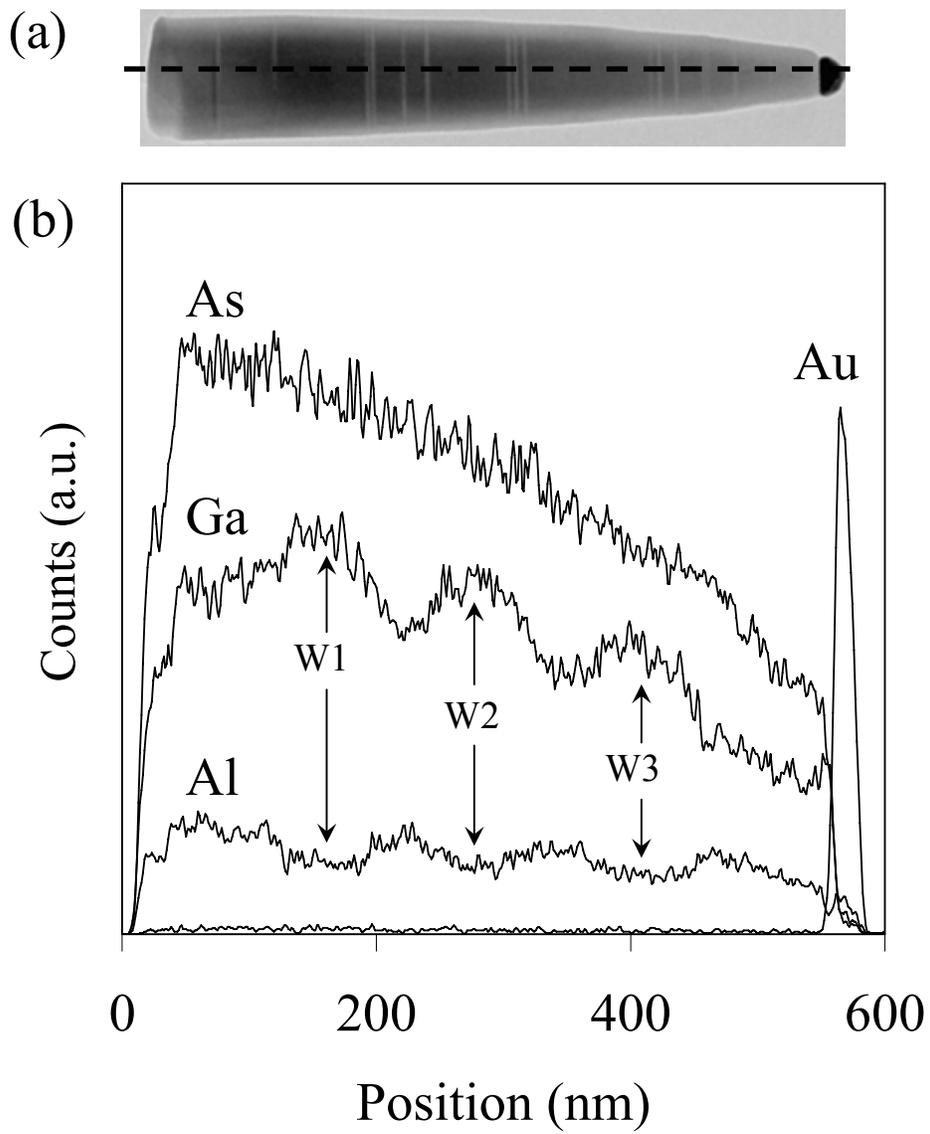

Figure 3

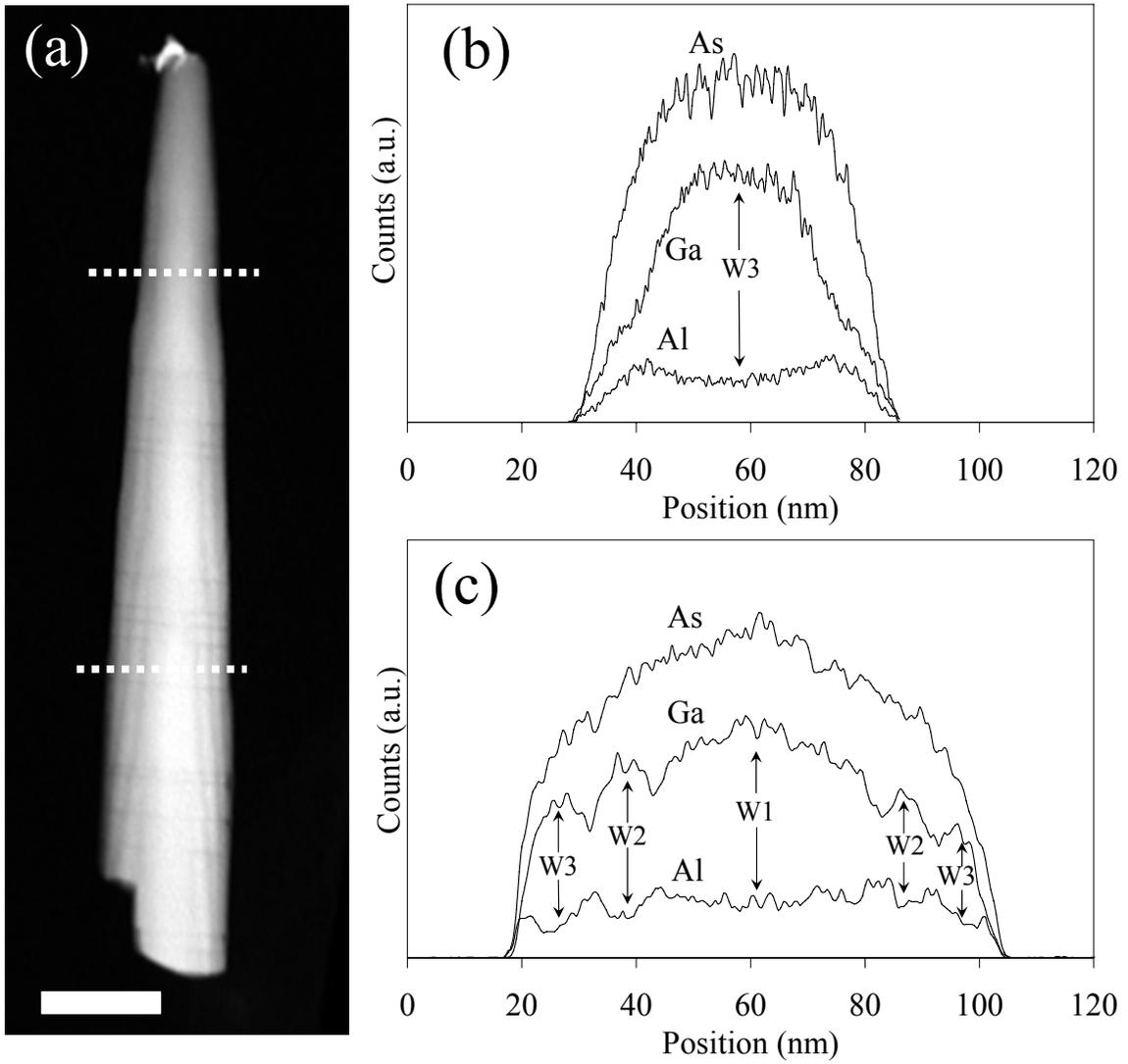

Figure 4

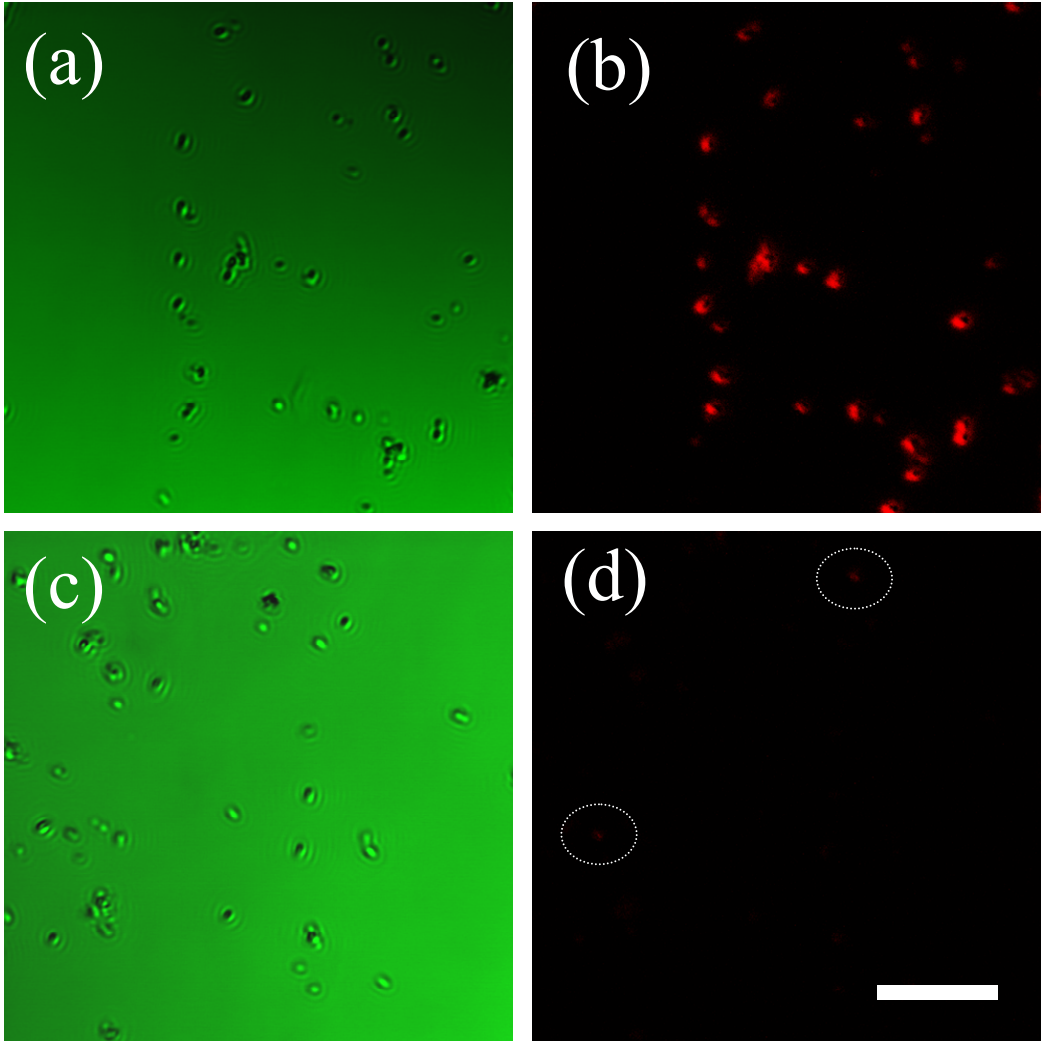

Figure 5

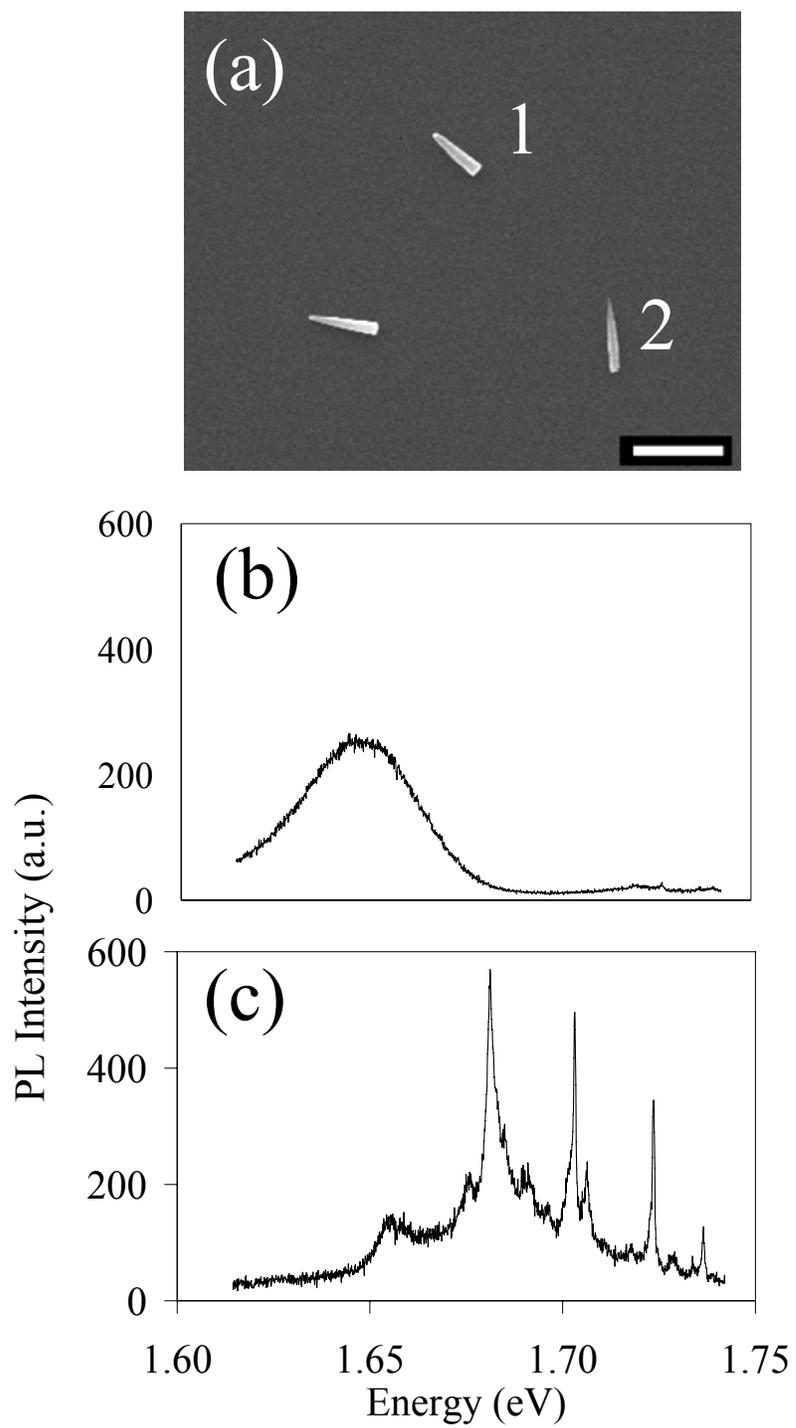